\documentclass[sigconf, nonacm]{acmart}
\usepackage{balance}

\AtBeginDocument{%
  }

\setcopyright{acmlicensed}
\copyrightyear{2026}
\acmYear{2026}
\setcopyright{cc}
\setcctype{by}
\acmConference[WSDM '26]{Proceedings of the Nineteenth ACM International Conference on Web Search and Data Mining}{February 22--26, 2026}{Boise, ID, USA}
\acmBooktitle{Proceedings of the Nineteenth ACM International Conference on Web Search and Data Mining (WSDM '26), February 22--26, 2026, Boise, ID, USA}
\acmPrice{}
\acmDOI{10.1145/3773966.3779404}
\acmISBN{979-8-4007-2292-9/2026/02}

\begin{document}

\title{On Listwise Reranking for Corpus Feedback}

\author{Soyoung Yoon}
\email{soyoung.yoon@snu.ac.kr}
\orcid{0009-0004-8669-8741}
\affiliation{%
  \institution{Seoul National University}
  \city{Seoul}
  \country{South Korea}
}

\author{Jongho Kim}
\email{jongh97@snu.ac.kr}
\orcid{0000-0003-2709-496X}
\affiliation{%
  \institution{Seoul National University}
  \city{Seoul}
  \country{South Korea}
}

\author{Daeyong Kwon}
\email{gwondaeyong63@gmail.com}
\orcid{0009-0003-1351-4625}
\affiliation{%
  \institution{Seoul National University}
  \city{Seoul}
  \country{South Korea}
}

\author{Avishek Anand}
\orcid{0000-0002-0163-0739}
\email{Avishek.Anand@tudelft.nl}
\affiliation{%
 \institution{Delft University of Technology}
 \city{Delft}
 \country{Netherlands}}

\author{Seung-won Hwang}
\orcid{0000-0003-0782-0661}
\email{seungwonh@snu.ac.kr}
\authornote{Corresponding author}
\affiliation{%
  \institution{Seoul National University}
  \city{Seoul}
  \country{South Korea}
}


\renewcommand{\shortauthors}{Soyoung Yoon, Jongho Kim, Daeyong Kwon, Avishek Anand, and Seungwon Hwang}

\begin{abstract}
Reranker improves retrieval performance
by capturing document interactions. At one extreme, graph-aware adaptive retrieval (GAR) represents an information-rich regime, requiring a pre-computed document similarity graph in reranking.
However, as such graphs are often unavailable, or incur quadratic memory costs even when available, graph-free rerankers leverage large language model (LLM) calls to achieve competitive performance. We introduce L2G, a novel framework that implicitly induces document graphs from listwise reranker logs. By converting reranker signals into a graph structure, L2G enables scalable graph-based retrieval without the overhead of explicit graph computation.
Results on the TREC-DL and BEIR subset show that L2G matches the effectiveness of oracle-based graph methods, while incurring zero additional LLM calls.\footnote{\url{https://github.com/soyoung97/l2g-submit}}
\end{abstract}

\ccsdesc[500]{Information systems~Information retrieval}
\ccsdesc[300]{Information systems~Retrieval models and ranking}
\ccsdesc[100]{Information systems~Language models}
\ccsdesc[100]{Information systems~Learning to rank}
\keywords{listwise reranker, adaptive retrieval, corpus graph}


\maketitle

\begin{figure}[t!]
{
\centering
    \includegraphics[width=0.95\linewidth]{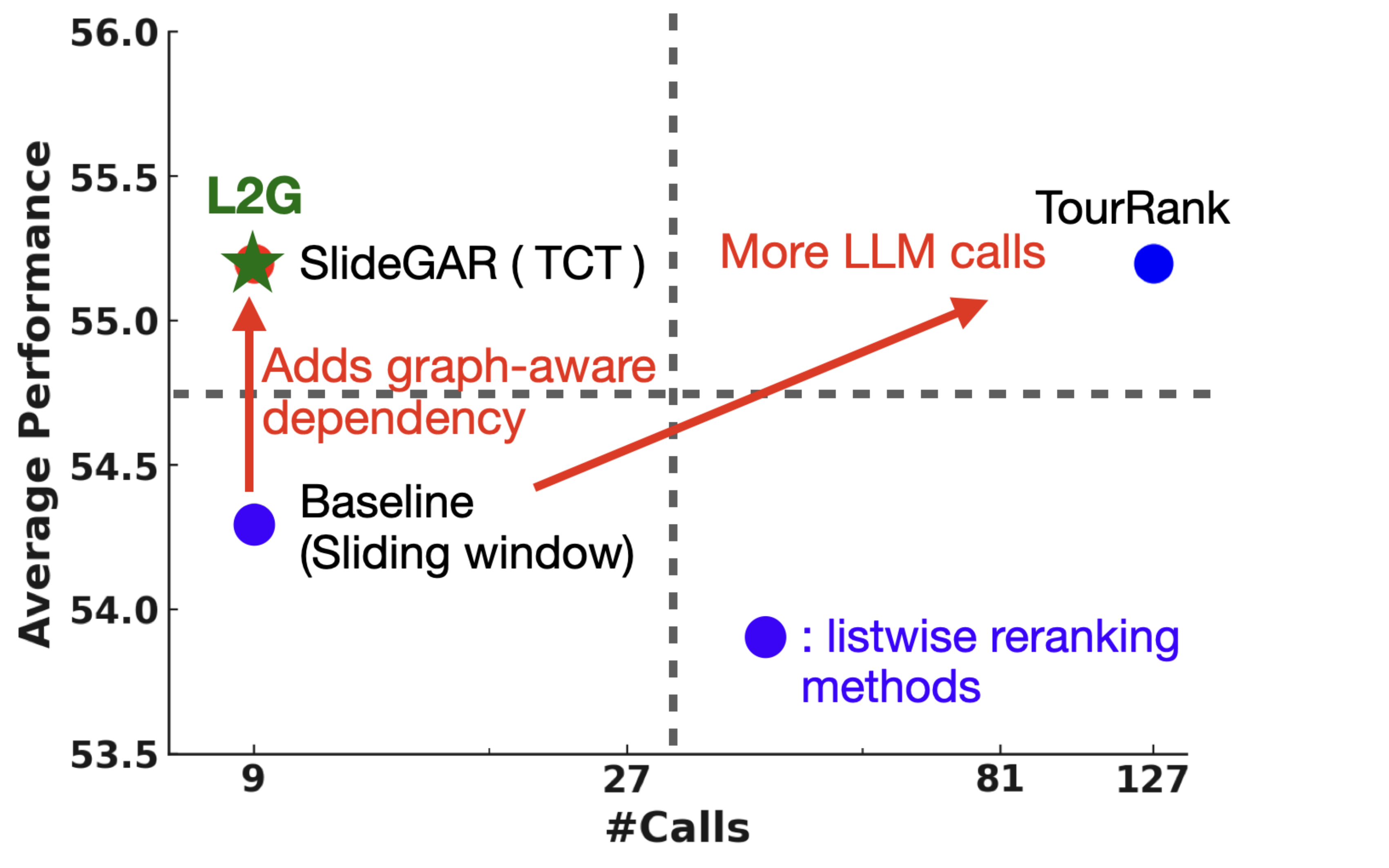}
    \caption{A cost-performance trade-off of rerankers 
    (nDCG@10; see Table \ref{tab:tourrank_acurank} for full results).  }

    \label{fig:motivation}
}

\end{figure}


\section{Introduction}

Reranking captures document interactions
as a form of corpus feedback to improve
retrieval performance~\cite{mao-etal-2024-rafe,rankzephyr,slidegar}.
To illustrate, 
listwise rerankers reflect local interactions by reranking documents within a window.
For a large input, this window shifts forward to compare high-ranking documents with the next unseen documents, until the window reaches the other end of the list. 
 However, such baselines, due to serialized window shifts,
fail to capture interaction feedback across non-adjacent windows.

One extreme of enriching corpus feedback is SlideGAR~\cite{slidegar}, which assumes a graph capturing all pairwise document interactions, then leverages the neighborhood information beyond the window from the graphs.
While effective, this approach requires full corpus graph access to construct them, and incurs substantial cost from repeated bi-encoder calls (e.g., TCT~\cite{lin-etal-2021-tctcolbert}), as well as a prohibitive $\mathcal{O}(N^2)$ memory footprint for a corpus of size $N$. In another extreme, graph-free rerankers such as TourRank~\cite{tourrank}  conduct multiple rounds of tournaments across windows.  Despite delivering a strong performance with SlideGAR without requiring explicit graphs, TourRank incurs heavy computation due to the large number of LLM calls required to remain competitive. 
 
Figure~\ref{fig:motivation} summarizes this tradeoff landscape by plotting average effectiveness against LLM calls. SlideGAR lies in the upper‑left: strong performance with sliding‑window baseline budget (9 calls), but requires additional compute and memory to construct the corpus graph. Another extreme in the upper-right is tournament‑style listwise reranker (e.g., TourRank) that is graph-free, improves performance but requires far more calls (about 127).

Our contribution is achieving high performance at low call budgets (upper‑left quadrant). Observing that listwise methods already expose local relations, we ask: can we reuse these signals across queries to approximate a document graph, without any additional LLM calls or requiring a doc-doc graph or retriever?

This motivates us to propose \textbf{L}istwise-to-\textbf{G}raph (\textbf{L2G}), requiring \emph{no} additional LLM or retriever calls while restoring SlideGAR performance. 
Our contributions are threefold: (i) we provide a yet effective perspective that frames listwise reranking as a source of implicit corpus feedback; 
(ii) we introduce \textbf{L2G}, which reconstructs an explicit doc-doc graph directly from ranking outputs (no separate doc–doc retriever); and 
(iii) we demonstrate SlideGAR performance with lower cost, making graph-free adaptive retrieval practical even in dynamic or resource‑constrained settings. 

\section{Related Works}

\textbf{Graph-based reranker}
Graph-aware methods such as \textbf{SlideGAR} \cite{slidegar} exploit \emph{explicit} doc-doc relations to guide adaptive reranking. While they achieve strong quality, they requiring pre-built corpus graphs with huge memory and pre-processing costs. Our work keeps the call budget low without requiring a doc-doc retriever by \emph{reconstructing} a sparse interaction graph directly from past listwise rankings and reusing it in SlideGAR.

\noindent \textbf{Graph-free reranker}
However, a corpus or graph is not always available. Then, advanced reranking for ranked results can be used to overcome the gap, albeit at an additional cost~\cite{rathee2025testtimecorpusfeedbackretrieval}.
We focus on
listwise rerankers~\cite{rankzephyr} that jointly reason over candidate sets. They often rely on sliding windows or tournament-style comparisons. RankZephyr~\cite{rankzephyr} is a strong fine-tuned listwise ranker; 
ReaRank~\cite{rearank} emphasizes reasoning ability with more inference time; TourRank~\cite{tourrank} improves effectiveness at the cost of more LLM calls. Contemporary advanced ranking approaches~\cite{coranking, demorank, slidingwindowsnotend} optimize in ICL/demo selection, collaborate small/large rankers with order adjustment, or expand context windows. 
Our distinction is that we frame listwise methods as \emph{implicitly} inducing local interaction graphs among candidates with zero LLM-call overhead.

\noindent \textbf{Query-stream reuse}
Transforming an entire corpus into a graph is a costly process, while previous work ~\cite{chen2025dontneedprebuiltgraphs} argued is unnecessary.
Our work shares this motivation, leveraging batch and stream processing~\cite{Ding2011batchqueryprocessing,eslami2019psidb,eslami2020query,mackenzie2023index} to demonstrate gains from shared computation and intermediate results across queries. Our incremental L2G updates operationalize this principle for reranking: listwise signals accumulate into a sparse graph that stays effective under order perturbations. 
\section{Method: Listwise signals to Graph (L2G)}
\label{sec:d2d_from_reranking}
\begin{figure}[t!]
\centering
\includegraphics[width=0.8\linewidth]{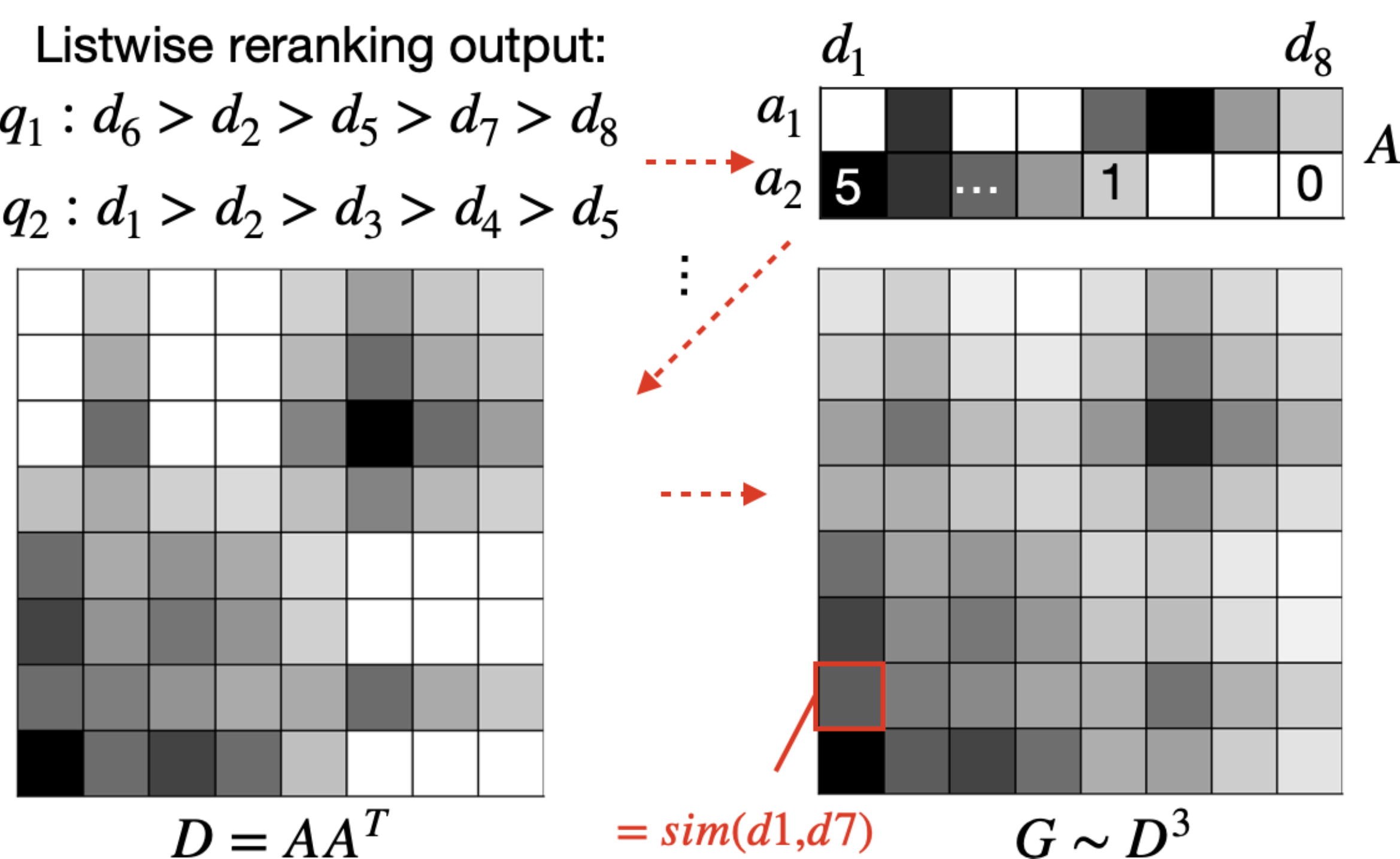}
\caption{Illustration of building $G$, a proxy of doc--doc graph, from listwise signals for L2G. Darker cells indicate higher pairwise relevance, and vice versa.}
\vspace{-5mm}
\label{fig:illustration}
\end{figure}

We study treating reranker outputs as document relationship signals and aggregate document pairs and their co-occurrence with queries,
which induces a first-order similarity matrix that serves as a proxy for an implicit graph over documents.
Fig.~\ref{fig:illustration} visualizes the building process. Let $\mathcal{Q}=\{q_{1},\dots,q_{m}\}$ be a stream of user queries.
For each $q_{i}$ a listwise reranker receives a candidate set
$\mathcal{C}_{i}=\{d_{i,1},\dots,d_{i,k}\}$ and returns a total
order $\pi_{i}:\mathcal{C}_{i}\!\to\![k]$,
where $\pi_{i}(d)$ is the rank of document $d$ for
query $q_{i}$ (one-dimensional).
We denote by $\mathcal{D}=\bigcup_{i}\mathcal{C}_{i}$ the (ever‑growing)
global document pool. 

\subsection{From Rankings to a First-order Graph}
\label{ssec:vectorizing}

First, we introduce a method for converting each ranked list for $q_i$ into a score vector. When accumulated, this forms a vectorized matrix for multiple queries, which becomes the basis for the underlying doc-doc graph structures. We map each ranked list into a dense \emph{document score vector} $\mathbf{a}_{i}\in\mathbb{R}^{|\mathcal{D}|}$ with $[\mathbf{a}_{i}]_{d}=(k-\pi_{i}(d)+1)$ if $d\in\mathcal{C}_{i}$, and $0$ otherwise. In other words, the scores get assigned from highest to lowest depending on their rank for each query. We approximate pairwise document affinity as $\mathbf{D}^{(1)}=\mathbf{A}\mathbf{A}^{\top}$, where $D^{(1)}_{d_{1},d_{2}}=\sum_{i}[\mathbf{a}_{i}]_{d_{1}}\,[\mathbf{a}_{i}]_{d_{2}}$ accumulates evidence over all queries in which \emph{both} documents co-occur, and treat this $\mathbf{D}^{(1)}$ as an approximation of the pairwise document similarity matrix. Related works suggest that greater query diversity leads to more accurate approximations~\cite{nystrom, inducingpoints}.

\subsection{Handling Sparsity and Higher‐Order Connectivity}
\label{ssec:n_hop}

The first-order similarity $\mathbf{D}^{(1)}$ is often sparse due to the limited overlap between candidate sets across queries. However, multi-hop connections—e.g., two documents may not co-occur but may each co-occur with a shared third document—can capture higher-order similarity structure. Formally, we extend the first-order graph by propagating affinity via $k$-hop random walks: $\mathbf{D}^{(k)}=\mathbf{D}\mathbf{D}\cdots\mathbf{D}$ ($k$ times), analogous to power iteration in PageRank~\cite{PageRank1999ThePC}. Larger $k$ increases recall but risks \emph{rank collapse} towards a uniform stationary distribution. We therefore (i) renormalize rows to unit $k_{1}$ norm after each multiplication and (ii) cap $k\le3$ in practice.

\noindent \textbf{Query-conditioned locality for robustness.}
While our L2G graph is defined over the full corpus, its induced edges are intentionally \emph{sparse} compared to fully precomputed pairwise graphs. To convert this sparsity into a precision advantage and avoid spurious hub expansion, L2G queries the graph \emph{locally}: relevant candidates are restricted to the first-stage retriever’s \texttt{top-$c$} pool (we use $c\in\{100,1000\}$). This query-conditioned locality reduces sensitivity to noisy long-range links when applying $k$-hop propagation, and improves by constraining exploration to a calibrated, high-precision pool.

\subsection{Handling Frequency Bias}
Still, the similarity scores in the matrix $\mathbf{D}$ can be skewed by document popularity rather than true semantic relevance. 
For example, documents that appear frequently across many queries will accumulate high similarity scores simply due to repeated co-occurrence, not because they are genuinely related.
Therefore, we borrow a well-established idea from information retrieval: \textbf{inverse document frequency (IDF) re-weighting}. For each document $d$, we divide its score vector by $\log\bigl(1+\mathrm{df}(d)\bigr)$, where $\mathrm{df}$ counts the number of queries that retrieved document $d$.

\subsection{Online Updates for Unseen Documents}
\label{ssec:incremental}

The challenge in maintaining the document matrix is that in practice, the document pool grows dynamically as new queries introduce previously unseen documents, making it impractical to recompute the entire similarity matrix $\mathbf{D}^{(1)}$ from scratch each time.
To support pairwise similarity for uncovered or newly added documents,
As a solution, we develop an incremental update strategy that efficiently extends the graph structure without full recomputation. Formally, the document pool approximates the document universe unknown \textit{a priori}. As a dynamic approximation, when a new batch of queries adds a set $\Delta\mathcal{D}$ of unseen documents, we update $\mathbf{A}_{\text{new}}=\begin{bmatrix}\mathbf{A}\\ \mathbf{A}_{\Delta}\end{bmatrix}$ and maintain $\mathbf{D}^{(1)}$ in block form instead of re-computing from scratch: $\mathbf{D}^{(1)}_{\text{new}}=\begin{bmatrix}\mathbf{D}^{(1)} & \mathbf{B}\\ \mathbf{B}^{\!\top} & \mathbf{C}\end{bmatrix}$, where $\mathbf{B}=\mathbf{A}\mathbf{A}_{\Delta}^{\top}$ and $\mathbf{C}=\mathbf{A}_{\Delta}\mathbf{A}_{\Delta}^{\top}$. This costs $\mathcal{O}(|\mathcal{D}|\,|\Delta\mathcal{D}|)$ rather than $\mathcal{O}(|\mathcal{D}_{\text{new}}|^{2})$. Since $\Delta|\mathcal{D}| \ll |\mathcal{D}|$ in typical streaming scenarios, this provides substantial computational savings.

\subsection{Complexity and Memory Footprint}
\label{ssec:complexity}
A key advantage of L2G is its elimination of expensive doc-doc similarity computations that plague existing graph-based methods. Unlike approaches such as SlideGAR-TCT that require explicit bi-encoder calls for every document pair, L2G introduces no additional LLM calls for doc-doc similarity and downstream modules simply read the needed relations directly from the on-the-fly graph G.

Operationally, the method requires expansion of $D$ upon the arrival of a new ranked list, and applies three hops of propagation to form $\mathbf{D}^{(3)}$ when a new query’s ranking arrives.  
In terms of space, L2G stores only the sparse graph \emph{G} (plus minimal ID maps), never materializing doc–doc matrices or document embeddings while entirely avoiding the need for any GPU-dependent doc-doc retriever.
Direct comparison of L2G over \texttt{SlideGAR--TCT} under a matched setup is further discussed at Sec.~\ref{sec:rq2}.
\section{Results}
\label{sec:results}
\begin{table*}[t]
\centering
\footnotesize
\scalebox{1.0}{%
\begin{tabular}{lcccccccccccccc}
\toprule
\multicolumn{1}{c}{} 
& \multicolumn{7}{c}{\textbf{BM25 Top-100}} 
& \multicolumn{7}{c}{\textbf{BM25 Top-1000}} \\
\cmidrule(lr){2-8}\cmidrule(lr){9-15}
& Covid & Sig. & News & Tou. & DL’19 & DL’20 & Avg
& Covid & Sig. & News & Tou. & DL’19 & DL’20 & Avg \\
\cmidrule(lr){2-8}\cmidrule(lr){9-15}
\multicolumn{1}{l}{Query count} 
& 50 & 97 & 57 & 49 & 43 & 54 & --
& 50 & 97 & 57 & 49 & 43 & 54 & -- \\
\multicolumn{1}{l}{Top-$c$ Overlap (\%)} 
& 10.4 & 0.8 & 3.3 & 1.3 & 0.1 & 0.1 & --
& 39.7 & 3.1 & 13.6 & 15.3 & 0.5 & 3.1 & -- \\
\multicolumn{1}{l}{Total corpus size} 
& 171k & 2.9M & 595k & 383k & 8.8M & 8.8M & --
& 171k & 2.9M & 595k & 383k & 8.8M & 8.8M & -- \\
\midrule
\multicolumn{15}{l}{\textit{(0) baseline — without doc2doc graphs}}\\
BM25 
& 59.5 & 33.0 & 39.5 & 44.2 & 50.6 & 48.0 & 45.4
& 59.5 & 33.0 & 39.5 & 44.2 & 50.6 & 48.0 & 45.4 \\
Sliding Window
& 84.1 & 32.0 & 52.3 & 32.4 & 74.0 & 70.2 & 57.5
& 80.7 & 28.9 & 51.0 & 30.9 & 75.1 & 78.8 & 57.6 \\
\midrule
\multicolumn{15}{l}{\textit{(1) with doc2doc graphs}}\\
SlideGAR--TCT (full corpus) 
& 80.1 & 31.0 & 51.7 & 34.9 & 74.2 & 79.3 & \textbf{58.5}
& 86.2 & 28.9 & 51.1 & 34.4 & 75.4 & 79.8 & \textbf{59.3} \\
SlideGAR--TCT (doc. affinity) 
& 83.0 & 31.1 & 53.0 & 37.5 & 74.2 & 71.1 & 58.3
& 85.0 & 29.5 & 52.7 & 32.7 & 75.1 & 80.0 & 59.2 \\
SlideGAR--Random
& 82.5 & 31.2 & 54.1 & 35.4 & 72.7 & 71.5 & 57.9
& 83.8 & 30.3 & 54.6 & 36.4 & 73.4 & 76.3 & 59.1 \\
\midrule
\textbf{L2G } 
& 84.2 & 31.8 & 53.4 & 37.4 & 72.3 & 71.2 & \textbf{58.4}
& 85.4 & 29.3 & 55.8 & 35.2 & 74.7 & 76.6 & \textbf{59.5} \\
\bottomrule
\end{tabular}}
\caption{Dataset stats and \textbf{NDCG@10 (\%)} with BM25 first-stage under two pool sizes (\textbf{Top-100} vs.\ \textbf{Top-1000}), using the RankZephyr-7B model. Benchmarks indicate TREC-Covid, Signal, News, Touche, TREC-DL19, and DL20, respectively. All variants use the same number of LLM calls as the Sliding Window.}
\vspace{-3mm}
\label{tab:rq1}
\end{table*}
Our empirical study addresses three questions. 
\textbf{1. Effectiveness}: can L2G, built only from past listwise rankings, 
recover doc-doc structure closely enough to match graph-aware oracles such as doc-doc affinity and full-corpus SlideGAR?
\textbf{2. Efficiency and cost}: does L2G keep the number of LLM calls fixed at the sliding-window budget while offering favorable latency and memory scaling compared to graph-based baselines?
\textbf{3. Robustness and generality}: is L2G stable under variations in query order and coverage, and does it transfer across different rerankers and first-stage retrievers?
\begin{table}[t]
\centering
\small
\setlength{\tabcolsep}{8pt}
\resizebox{0.8\linewidth}{!}{%
\begin{tabular}{lcc}
\toprule
 & \textbf{\begin{tabular}[c]{@{}c@{}}SlideGAR-TCT \\ (doc. affinity)\end{tabular}} & \textbf{L2G (ours)} \\ \midrule
Extra store & \begin{tabular}[c]{@{}c@{}}doc--doc \\ embeddings \end{tabular} & \begin{tabular}[c]{@{}c@{}}No pre-built \\ embeddings\end{tabular} \\  \midrule
Latency / query (s) & 0.427 & \textbf{0.103} \\ 
Peak Storage (MB) & 0.862 & \textbf{0.855} \\ 
Peak VRAM (MB) & 418.73 + $\alpha$ & None \\ \bottomrule
\end{tabular}
}
\caption{Efficiency comparison between SlideGAR-TCT and L2G, averaged for all reported BEIR subsets. Storage is measured as the in-memory footprint of the data structure (in bytes). Smaller is better; L2G achieves consistently lower latency and memory without prebuilt corpus embeddings.}
\vspace{-3mm}
\label{tab:efficiency}
\end{table}

\subsection{Experimental Setup}
\textbf{Datasets \& Evaluation.}
We evaluate on MS~MARCO TREC~\cite{craswell2020overview} DL’19/’20 (43/54 queries) and a subset of BEIR~\cite{beir} collections (TREC-COVID, Signal, TREC-NEWS, and Touche) with query size less than 100, and report \textit{nDCG@10}~\cite{ndcg} under reranking budget $c$ (c= 100 or 1000) rounded to the nearest tenth, mostly prepared with the pre-built Pyserini index~\cite{pyserini}.

\noindent \textbf{Retrieval \& Ranking Models.}
We use BM25~\cite{bm25} and extend to Contriever~\cite{contriever} as the first-stage retriever. 
Listwise rerankers include \textit{RankZephyr}~\cite{rankzephyr} andtextit{ReaRank}~\cite{rearank}, operating with sliding windows (size=20, step=10) that output permutations without explicit scores.

\noindent \textbf{Baselines.}
We compare to: (i) non-adaptive \textit{Sliding Window}, and 3 different (ii) Graph-based \textit{SlideGAR--TCT} baselines. (1) \textit{full-corpus} assumes oracle access to a pre-computed TCT~\cite{tct} graph over the entire collection. (2) \textit{doc–doc affinity} is a more practical baseline, which mimics the online graph construction scenario just like ours (Sec.~\ref{ssec:n_hop}), limiting the graph neighbors to a \emph{local} graph built only from the query’s retrieved top-$c$ candidates, e.g., top-100. Lastly, (iii) \textit{Random} randomizes neighbor ordering from local top-$c$ candidates, verifying that L2G's gains are not due to chance.


\subsection{Effectiveness}
\label{sec:rq1}
Table \ref{tab:rq1} summarizes nDCG@10 on MS MARCO DL’19/’20 and BEIR subsets with BM25 first-stage pools ($c$ $\in$ 100, 1000). First, we observe that L2G consistently outperforms the sanity-check baseline of a random doc-doc ranker (SlideGAR-Random) across all setups. Under Top‑100, L2G effectively ties the graph‑based oracles—SlideGAR–TCT with doc–doc affinity and the full‑corpus variant—within $\approx$ 0.1 nDCG on average, all at the same LLM‑call budget as the sliding‑window reranker. Moving to Top‑1000, L2G even \textit{wins} over full‑corpus SlideGAR–TCT while relying only on local interaction signals. We hypothesize this increased effectiveness due to the larger top‑c overlap reported in the table, which lets L2G reuse cross‑query evidence to enrich its induced graph. This indicates that once overlap is sufficient, listwise‑induced graphs recover—and in some cases exceed—the benefits of explicit corpus graphs without building them.

\begin{table}[t]
\centering
\small
\setlength{\tabcolsep}{4pt}
\resizebox{0.9\linewidth}{!}{
\begin{tabular}{lccccc}
\toprule
Method / Setting & Covid & Sig. & News & Tou. & Avg \\
\midrule
\multicolumn{6}{l}{\textbf{(1) Perturbing query stream} (BM25 top-100, RankZephyr)} \\
\midrule
L2G (dataset order) & 84.2 & 31.8 & 53.4 & 37.4 & \textbf{51.7} \\
L2G (max overlap)    & 83.0 & 31.4 & 53.5 & 37.6 & 51.4 \\
L2G (min overlap)    & 83.8 & 31.7 & 53.9 & 36.8 & 51.6 \\ \midrule
\multicolumn{6}{l}{\textbf{(2) Different first-stage retriever} (Contriever top-100, RankZephyr)} \\
\midrule
Sliding Window            & 70.8 & 26.5 & 49.9 & 30.4 & 44.4 \\
SlideGAR (doc. affinity)  & 72.1 & 25.0 & 51.9 & 29.7 & \textbf{44.7} \\
SlideGAR-Random         & 72.0 & 25.0 & 50.1 & 27.1 & 43.6 \\
L2G                       & 72.0 & 26.8 & 50.9 & 28.9 & \textbf{44.7} \\ \midrule
\multicolumn{6}{l}{\textbf{(3) Different reranking model} (Contriever top-100, ReaRank)} \\
\midrule
Sliding Window            & 71.6 & 27.0 & 50.5 & 24.4 & 43.4 \\
SlideGAR (doc. affinity)  & 71.6 & 26.9 & 50.4 & 23.9 & 43.2 \\
SlideGAR-Random        & 74.4 & 27.2 & 51.3 & 23.6 & 44.1 \\
L2G                       & 74.3 & 27.3 & 52.4 & 24.7 & \textbf{44.7} \\
\bottomrule
\end{tabular}}
\caption{RQ3: Robustness \& Generalizability (NDCG@10, \%).}
\vspace{-3mm}
\label{tab:rq3}
\end{table}
\subsection{Efficiency / Cost}
\label{sec:rq2}
We evaluate only the adaptive stage (excluding first‑stage retrieval and any LLM calls), reporting per‑query latency and peak memory. For L2G, \textbf{latency} is measured from the moment a ranked list for query $q$ arrives: we measure the per-query cumulative time to expand the document pool, form $D{=}AA^\top$, and yield $D^3$. We compare L2G with \texttt{SlideGAR–TCT (doc‑affinity)}, as it is a strong baseline that also constructs a graph online. 
As summarized in Tab.~\ref{tab:efficiency}, SlideGAR–TCT averages $0.427$\,s per-query, whereas L2G requires no precomputation and remains well below this even as the maintained graph grows. On \textbf{memory}, SlideGAR–TCT must load a bi‑encoder (about $419$\,MB), store top‑100 doc-doc affinity graph dense embeddings (once per query in the first window), and maintain a neighbor dictionary (doc IDs with float scores).
L2G avoids the bi‑encoder and dense vectors entirely: its footprint is a single sparse graph over documents actually observed along with the minimal docID maps, scaling with unique candidate documents. Full‑corpus graphs are not comparable on a per‑query basis but are much heavier than the online ones; e.g., for DL’19 the full‑corpus TCT dump occupies $\sim$3.1\,GB on disk, and even the smaller Touche collection takes $>$1\,hour to dump on our machine, while many scenarios do not even provide full‑corpus access. Taken together, L2G delivers doc‑affinity–level effectiveness at a fraction of the runtime and with a smaller footprint. In practice, delaying or batching graph updates (e.g., delay three propagation hops) further reduces constant factors without affecting ranking quality, which we leave as future work. 

\subsection{Robustness / Generalizability}

\textbf{Ordering robustness.}
While document overlap between queries is crucial for L2G's graph construction as discussed in Sec.~\ref{sec:rq1}, a potential concern with L2G's streaming approach is sensitivity to query arrival orders. We address this concern by analyzing L2G's robustness to query order variations within the same benchmark.
(i) a \textbf{max overlap} order greedily picks the next query whose top‑100 overlaps the most with the pool already processed, and (ii) a \textbf{min overlap} order greedily minimizes this cumulative reuse.
Tab.~\ref{tab:rq3}–(1) further shows that what matters is the overall amount of overlap present in the corpus, not the \textit{specific order} in which overlapping queries arrive; although Table~\ref{tab:rq1} suggested that effectiveness increases with higher corpus-level overlap, results confirm that the arrival order itself is \textit{not} directly related to output performance. 

\noindent\textbf{Generalizability.} On Tab.~\ref{tab:rq3}--(2) and (3), we observe that L2G dominates over SlideGAR-TCT (doc-affinity) even when swapping first-stage retrievers (BM25 $\rightarrow$ Contriever) and base listwise rankers (RankZephyr $\rightarrow$ ReaRank). Overall, L2G behaves as a reusable graph prior that transfers across ranking stacks.

\section{Conclusion}

This work explores listwise rerankers as an alternative to graph-based adaptive retrieval when a graph is unavailable or costly to build. 
We demonstrate how the graph structure can be constructed from the listwise reranking results and reused across queries. We validate that our L2G performs on par with graph-based oracles.
We are first to frame listwise reranking as an alternative form of adaptive retrieval, bridging the gap between graph-based and graph-free reranking.


\begin{acks}
This work was partly supported by Institute of Information \& communications Technology Planning \& Evaluation (IITP) grant funded by the Korea government(MSIT) [NO.RS-2021-II211343, Artificial Intelligence Graduate School Program (Seoul National University)], 
and the National Research Foundation of Korea(NRF) grant funded by the Korea government(MSIT) (No. RS-2024-00414981), 
\end{acks}

\newpage
\section{Ethical Considerations}

While the L2G is far more efficient than SlideGAR, additional requirements for graph construction may still limit accessibility for resource-constrained organizations.
The efficiency of L2G can be further optimized by tuning graph update intervals at the cost of performance, or through implementation of the GPU version of L2G. Our simple implementation leaves additional speed and memory gains as future work, which could help address these accessibility concerns.

\bibliographystyle{ACM-Reference-Format}
\balance
\bibliography{custom}

\appendix
\newpage
\section*{Appendix}
\begin{table*}[t]
\centering
\resizebox{\linewidth}{!}{
\begin{tabular}{l
                c c c c c c
                c c c c c c
                c c c c c
                c}
\toprule
\textbf{Doc Ranker} & \textbf{D19} & \textbf{D20} & \textbf{D21} & \textbf{D22} & \textbf{D23} & \textbf{D-H} &
\textbf{AVG} & \textbf{covid} & \textbf{news} & \textbf{nfc} & \textbf{sig.} &
\textbf{r04} & \textbf{touc.} & \textbf{dbp} & \textbf{SciF.} & \textbf{\textbf{BEIR}} & \textbf{\textbf{ALL}} & \textbf{\#Calls} \\ \midrule
BM25              & 50.6 & 48.0 & 44.6 & 26.9 & 26.2 & 30.4 &
\textbf{37.8} & 59.5 & 39.5 & 32.2 & 33.0 &
40.7 & 44.2 & 31.8 & 67.9 & \textbf{43.6} & \textbf{41.1} & -- \\[1pt]
Sliding Windows    & 74.0 & 70.2 & 69.5 & 51.5 & 44.5 & 38.6 &
\textbf{58.0} & 84.1 & 52.3 & 36.8 & 32.0 &
54.0 & 32.4 & 44.5 & 75.5 & \textbf{51.5} & \textbf{54.3} & 9 \\[1pt]
SlideGAR (BM25) & 73.6 & 72.1 & 68.4 & 49.8 & 45.2 & 37.4 &
\textbf{57.8} & 82.9 & 53.8 & 36.9 & 31.6 &
54.6 & 37.0 & 43.7 & 75.5 & \textbf{52.0} & \textbf{54.5} & 9 \\[1pt]
SlideGAR (TCT)  & 74.2 & 72.3 & 68.9 & 51.1 & 46.7 & 39.5 &
\textbf{58.8} & 83.3 & 54.4 & 36.6 & 32.3 &
55.5 & 37.9 & 43.9 & 75.6 & \textbf{52.4} & \textbf{55.2} & 9 \\[1pt]
TourRank-1  & 74.2 & 68.2 & 69.6 & 51.1 & 45.2 & 38.1 & \textbf{57.7} & 81.8 & 36.5 & 30.7 & 51.9 & 54.5 & 31.2 & 43.2 & 71.3 & \textbf{50.1} & \textbf{53.4} & 12.7\\[1pt]
TourRank-3  & 73.8 & 70.1 & 71.2 & 52.3 & 47.0 & 40.7 & \textbf{59.2} & 83.0 & 37.3 & 31.7 & 51.9 & 56.4 & 33.5 & 44.4 & 74.2 & \textbf{51.5} & \textbf{54.8} & 38.2\\[1pt]
TourRank-10 & 74.9 & 71.8 & 71.4 & 53.3 & 47.7 & 39.9 & \textbf{59.9} & 83.2 & 37.1 & 31.1 & 53.3 & 57.1 & 32.1 & 44.8 & 75.1 & \textbf{51.7} & \textbf{55.2} & 127.4 \\[1pt]
\bottomrule
\end{tabular}}
\caption{Document–ranking performance (higher is better).  
SlideGAR rows are shaded gray. The \textbf{BEIR} and \textbf{ALL} columns show averages and are boldfaced for emphasis. SlideGAR uses \textbf{lowest} llm calls with competitive ndcg performance, but advanced listwise reranking methods like TourRank can keep up with SlideGAR with the cost of additional LLM calls.}
\label{tab:tourrank_acurank}
\end{table*}
Table~\ref{tab:tourrank_acurank} refers to the full results on the comparison between advanced listwise reranking methods v.s. slidegar (full corpus).

\section{Additional Discussion on the Efficiency of L2G}
Building on Sec.~\ref{sec:rq2}, we further discuss the efficiency of L2G, with a particular focus on how its \emph{session-induced} graph scales compared to baselines that rely on a \emph{corpus-level} document--document affinity graph (e.g., SlideGAR-TCT).

\paragraph{Setup and fairness.}
To keep the comparison conservative and even slightly favorable to the document--document affinity baseline (SlideGAR-TCT), we (i) enable \emph{document caching} and \emph{GPU batching} for TCT, and (ii) use 768-dimensional FP32 vectors for both queries and documents across all systems.
All measurements are obtained on a GPU server with dual Intel Xeon Gold 6254 CPUs (36 cores total; 72 threads), 607\,GB system memory, and 8 NVIDIA GeForce RTX 3090 GPUs (24\,GB each).
For reproducibility, the code used to obtain the efficiency results is available in the following GitHub branch.\footnote{\url{https://github.com/soyoung97/l2g-submit/tree/efficiency}}

\subsection{Compute with respect to document pool size.}
\paragraph{Compute (wall time).} Since the graph grows dynamically as new queries add signals from previously unseen documents, the computational cost of the graph propagation step ($G \sim D^3$) and online updates does grow as the document pool scales. With $k{=}20$ and eager graph updates ($U{=}1$), L2G shows near-linear per-query latency in the number of processed queries $q$:
\[
    \ell_{\mathrm{L2G}}(q) \approx 0.035\,q~\text{s}.
\]
However, we believe there is substantial room for improvement. This can be further optimized by using advanced caching techniques and efficient graph storage algorithms, such as pre-allocating document slots in batches to accommodate anticipated graph growth. Currently, the graph is updated whenever a single new ranking signals arrive; by batching or delaying graph updates (i.e., increasing the update interval $U$), the constant in the $0.035\,q$ slope would shrink proportionally, yielding faster runtime without changing the scoring model. We leave such optimizations as future work.

\paragraph{Memory.} Peak working memory for L2G also scales linearly with the number of processed queries $q$:
\[
    \mu_{\mathrm{L2G}}(q) \approx 0.0276\,q~\text{MB}.
\]
In contrast, the full-corpus doc-doc affinity variant of SlideGAR (without Top-100 truncation) incurs memory size and graph construction latency that scale tremendously with corpus size. A key advantage of L2G is that its resource consumption depends only on $q$, not on the full corpus size.

\paragraph{Takeaways.} (i) L2G attains SlideGAR-level quality without any offline graph building; (ii) incremental sparse updates reduce compute and memory versus naïve $|\mathcal{D}|^2$ construction and provide a tunable cost-latency knob ($U$) that is largely orthogonal to model quality; (iii) even in a SlideGAR-favored configuration (document caching and GPU batching enabled), L2G is faster and lighter for small to medium sessions, and remains competitive thereafter. Overall, \emph{cost parity with SW and quality parity with doc-doc affinity} make L2G an attractive default when storing a small sparse graph is acceptable.

\section{Limitations}

Our method's effectiveness is highly affected on documents co-occurring in retrieved lists, making it less effective in scenarios with low query-document overlap. In this case, when the constructed graph remains sparse, L2G may not provide substantial benefit over simpler baselines.

\end{document}